# Study of pinning at 3D-2D phase transition in vortex matter of strongly anisotropic high-temperature superconductors of BiPbSrCaCuO system.

S. M. Ashimov, J. G. Chigvinadze

Sharp increase of pinning force was observed at the 3D-2D phase transition in strongly anisotropic high temperature superconductors of the BiPbSrCaCuO system.

In work [1] published in 2000 it was shown that in definite conditions in layered high-temperature superconductors the disintegration of three-dimensional 3D Abrikosov vortices into "pancake" quasi-two-dimensional 2D ones could take place. This 3D-2D phase transition in vortex matter must be accompanied by a sharp increase of critical current.

A critical current increase should be naturally connected and supported by the increase of pinning force acting on a vortex matter. The 3D-2D phase transition apparently promotes an increase of local pinning force because disintegrating three-dimensional macroscopic vortex lines are transformed into quasi-two-dimensional macroscopic formations of vortex matter with one of their dimensions – thickness being microscopical in the investigated samples and about 1.4 Å, and by this smaller then atomic dimensions of high-temperature superconductor's material. Thus, following work [1], at the disintegration of volume vortices on "pancake" ones a sharp increase of pinning force should be observed. This conclusion on a possibility of pinning force increase at 3D-2D transition is supported by experimental works [2,3], where it was shown that at the study of field dependence of vibration damping $\delta$ in a mixed state at definite magnetic fields (of the order of 2000 Oe) it sharply decreased. The decrease of $\delta$ was so large that the damping at fields $H$ ~2000 Oe and higher turned to be equal the background one observed in the absence of magnetic field. This large change of damping is connected with 3D-2D phase transition in the vortex matter.

But for the final confirmation of the conclusions of work [5], a direct measurement of pinning force should be made because at increase of critical current at 3D-2D phase transition the pinning force should correspondingly increase. This work is namely devoted to the clearing up of this problem.

For the measurements of pinning force the specially elaborated mechanical (contactless) method [4] of the pinning force direct measurement in high-temperature superconductors is applied. The method is based on measurements of countermoments (counterforces), acting on axially symmetrical superconductive sample at its rotation in a transverse outer magnetic field.

The determination of countermoments of pinning forces and viscous friction acting on a sample from its system of quantum vortex lines (VL) was made accordingly the principle suggested in works [5,6]. Method's sensitivity is sufficiently high and as it was shown in work [7] it is equivalent to ~$10^{-8}$ Vcm$^{-1}$ in the method of investigations of current (versus) voltage characteristics.

Principal layout and experiment's geometry are shown in fig. 1. In the experiment the rotation angle $\varphi_2$ of sample is measured depending on the rotation angle $\varphi_1$ of rotating head setting a sample in the rotation by a suspence thread with the torsional stiffness $K \approx 4 \cdot 10^{-8}$ Nm which is substituted by the stiffer one when it is necessary. Investigation were carried out at a constant rotation velocity of twisting head $\omega = 1.8 \cdot 10^{-2}$ rad.sec$^{-1}$.

Rotation angles $\varphi_1$ and $\varphi_2$ were correspondingly defined with the accuracy of $\pm 4,6 \cdot 10^{-3}$ rad and $\pm 2,3 \cdot 10^{-3}$ rad. The homogeneity of magnetic field strength along a sample was $\Delta H / H = 10^{-3}$.

To avoid effects connected with frozen magnetic fluxes the lower part of cryostat with a sample was placed in a special permalloy shield reducing Earth magnetic field's level down in 1200 times. Then a sample was reset in superconductive state, the shield removed, a magnetic field of desired strength H applied and a dependence $\varphi_2(\varphi_1)$ taken. To carry out measurements at other value of H the sample was brought in the normal state by heating up to $T > T_c$ at $H = 0$, and only after the resetting of sample and twisting head in the initial state $\varphi_1 = \varphi_2 = 0$ the experimental procedure was repeated.

Last time investigations of vortex matter are carried out on monocrystal samples of high-temperature superconductors. The use of monocrystals as samples certainly makes the interpretation of obtained results easier. But due to tiny sizes of obtained high-temperature monocrystals (123) type the effects observed in them and connected with the vortex lattice dynamics or other investigations related with the study of interaction processes of vortex matter with crystal lattice of superconductors are so weak that their registration is sometimes difficult.

It is problematic also the fabrication of monocrystals with desired sizes and quality for other systems of high-temperature superconductors. On the base of above pointed we decided to make texturized samples of HTSC materials of large sizes which **c**-axes could be directed at one's desire as along a cylindrical sample used by us (in our case magnetic field H in directed perpendicular to the cylinder axes) and to study the vortex lattice dynamics in a base plane of HTSC, as well as to direct **c**-axes perpendicular to the cylinder axis and to investigate the vortex lattice dynamics along **c**-axes. Just this last situation was realized in the present experiments when magnetic field $H \| c$. Samples used by us were described in work [2].

Typical dependences $\varphi_2(\varphi_1)$ taken at T=95 K for $Bi_{1,7}Pb_{0,3}Sr_2Ca_2Cu_3O_{10-\delta}$ sample are presented in Fig.2. At the rotation of sample both in the normal and superconductive states in the absence of outer magnetic field ($H = 0$) the $\varphi_2(\varphi_1)$ dependence is straight (marked by a dotted line) and the condition $\varphi_1 = \varphi_2 = \omega t$ is fulfilled.

The character of $\varphi_2(\varphi_1)$ dependence is essentially changed when a sample is placed in magnetic fields $H > H_{c1}$ at $T < T_c$. In figure the curves taken in fields $H = 300$, 400 and 500 Oe are presented. Their typical character are revealed at measurements up to $H = 1500$ Oe and two regions are clearly seen in them. As it is seen from the figure sample's rate gradually increases as far as $\varphi_1 \sim \tau = K(\varphi_1 - \varphi_2)$ increases what results from a progressing process of VL's detached from the corresponding pinning centers. One should wait that just in this region in a rotating sample the "vortex fan" is opened in which VL are distributed along instant orientation angles relatively the fixed outer field. During this values of separate VL's orientation angles are limited by $\varphi_{fr}$ and $\varphi_{fr} + \varphi_{pin}$, where $\varphi_{fr}$ is an angle by which VL could be rotated relatively H by forces of viscosus friction with superconductor's matrix, but $\varphi_{pin}$ is an angle by which VL could be rotated by the most strong pinning center firstly studied in work [8].

The gradual transition of (at large values of $\varphi_1$) first range into the second one where the dependence $\varphi_2(\varphi_1)$ is observed makes it possible to define independently one from other the countermoments of pinning forces $\tau_p$ and viscous forces $\tau_{fr}$. Namely in this regime when $\omega_2 = \omega_1$, the twisting moment $\tau$ applied to a uniformly rotating sample is balanced by countermoments $\tau_p$ and $\tau_{fr}$.

In particular, in works [9,10] it was shown that in the uniformly rotating sample mode

$$\tau = \tau_p + \tau_0 \omega_2 ,$$

where

$$\tau_0 = \frac{\pi^2}{2} \eta B \Phi_0^{-1} L R^4 ,$$

here $B$ is the induction in a sample; $\Phi_0$ is the quantum of magnetic field; $\eta$ is the viscous friction factor; $L$ is the length and $R$ is the radius of sample.

If the twisting head is jammed in this range then a sample due to the relaxation processes related with viscous forces acting on VL would continue rotation in the same direction (with a decreasing rate) before the same equilibrium position could be obtained depending on the value of H [11].

Curves of time dependences $\Delta\varphi_2^{rel}$ of the stopped twisting head at T=95 K and $H$ =300, 400 and 500 Oe are presented in the insertion of Fig.2. By the angle of relaxation rotation $\Delta\varphi_2^{rel}$ one could easily define the countermoment of viscous forces $\tau_{fr}$ acting on VL during the rotation process:

$$\tau_{fr} = k \Delta\varphi_2^{rel} ,$$

Accordingly the expression (1), the twisting moment $\tau$ applied to a sample after its relaxation rotation and stopping ($\omega_2 = 0$) would be only balanced by the countermoment of static pinning forces

$$\tau_p = k \left[ \varphi_1 - (\varphi_2 + \Delta\varphi_2^{rel}) \right],$$

where $\varphi_1$ and $\varphi_2$ is values of angles in the stopping moment of twisting head.

Essentially different character of $\varphi_2(\varphi_1)$ dependences is observed in magnetic fields $H \geq 1600$ Oe. As example, in Fig.2 these dependences in fields $H$ =2000 Oe and $H$ =4000 Oe are presented. In this case a clear difference is seen from $\varphi_2(\varphi_1)$ dependences observed in the comparatively weak magnetic fields, namely, one could see three regions and the region $\omega_2 = \omega_1$ which is a characteristic of weak fields $H$ <600 Oe. is absent. The first (beginning) is the region where the sample doesn't respond to the twisting moment $\tau$ applied to it and increasing with time or responds comparatively weakly. Such behavior of sample could be explained by the fact that VL are not detached from pinning centers at small values of $\tau$ but an insignificant rotation of sample is caused by the elastic deformation of magnetic force lines out it, possibly, by a detachment the most weakly fixed VL. With the increase of $\varphi_1 \sim \tau$ beginning from some critical value of $\varphi_c^{min}$, depending on the strength of magnetic field H, The process of VL detachment and a gradual growth of $\varphi_2$ starts. This second region on the $\varphi_2(\varphi_1)$ dependence as it is seen from the figure is sufficiently prolonged and characterized by a continuous growth of twisting moment $\tau \propto \varphi_1$ applied to the sample.

The increase of $\varphi_1 \propto \tau$ up to the third region, where the process of VL detachment is essentially accelerated and after the reaching some $\varphi_c^{max}(H)$ the massive detachment of VL is observed from corresponding pinning centers, is accompanied by the stepwise increase of $\varphi_2$ and decrease of $\tau_p$ value.

As it was shown in work in work [10] the measurements of $\tau_p(H)$ dependences makes it possible to define also the pinning force

$$F_p = \frac{3}{4}\tau_p R^{-3} L^{-1}.$$

In Fig.3 the dependences of volume pinning force on a strength of magnetic field H are presented. As it is seen from the figure in weak fields the growth of average volume pinning force is observed passing through the maximum. This region is apparently related with volume 3D Abtikosov vortices and their pickup on pinning centers which in all probability exists in strongly anisotropic high-temperature superconductors. Starting from the field $H \approx 600$ Oe and higher $F_p$ increases linearly but with other slope being different from the one of the first region.

Further on at fields close to $H \approx 2000$ Oe a sharp stepwise growth of the pinning force is observed. In a narrow region of magnetic field $F_p$ increases almost on 300%. After this jump again a linear increase of $F_p$ follows, but also in this case the curve slope is changed. Because these experiments on the pinning investigation were carried out on the same strongly anisotropic high-temperature superconductors of BiPbSrCaCuO system where the dissipation processes were studied then analysis carried out in work [2] could be applied also to our case. It was shown there that accordingly an assessment the 3D-2D phase transition should take place at fields of the order of $H \approx 2000$ Oe.

Just at these fields the dissipation processes were sharply suppressed what made it possibly to conclude and relate observed effects with a possible 3D-2D phase transition.

Using the assessment of field value for these phase transition made in work [2] which is evaluated in [12] by the relation

$$B_{2D} \approx \frac{\Phi_0}{\Gamma d^2},$$

where $\Phi_0$ is the magnetic flux quantum $\Phi_0 = \frac{hc}{2e} = 2.07 \cdot 10^{-7}$ G cm$^2$, $\Gamma$ is the anisotropy factor, it turns to be approximately equal to $H \approx 2000$ Oe.

Taking into account that just in these region of magnetic fields the stepwise increase of pinning force take place, we could make statement that in strongly anisotropic high-temperature texturized superconductor of $Bi_{1.7}Pb_{0.3}Sr_2Ca_2Cu_3O_{10-\delta}$ in magnetic fields $H \approx 2000$ Oe apparently 3D-2D phase transition in the vortex matter is observed and the conclusions of theoretical work [1] on the possibility of phase transition in the vortex matter are thereby confirmed.

The work is made in the frames of ISTC G-389 Grant.

Figure captions:

Fig.1. Experimental layout. 1 – sample, 2 – upper elastic thread, 3 – lower thread, 4 – twisting head, 5 – straighted glass rod.

Fig.2. Dependence of sample rotation angle $\varphi_1$ on the twisting angle $\varphi_2$.

Fig.3. Dependence of average volume pinning force on the strength of outer magnetic field.